\documentclass[a4paper,fleqn,usenatbib]{mnras}

\usepackage{newtxtext,newtxmath}
\usepackage[T1]{fontenc}
\usepackage{ae,aecompl}

\usepackage{graphicx}	% Including figure files
\usepackage{amsmath}	% Advanced maths commands
\usepackage{amssymb,multirow}	% Extra maths symbols

\usepackage{bm}
\title[Accretion driven by HJs]{Pulsed Disc Accretion Driven by Hot Jupiters}
\author[Teyssandier \& Lai]{Jean Teyssandier$^{1,2}$\thanks{E-mail: jean.teyssandier@unamur.be} and Dong Lai$^{1}$\\ %$^{1}$
$^1$ Department of Astronomy, Cornell Center for Astrophysics and Planetary Science, Cornell University, Ithaca, NY 14853, USA \\
$^2$ naXys - Department of Mathematics, University of Namur, 8 Rempart de la Vierge,
5000 Namur, Belgium
}

\newcommand{\dd}[2]{\frac{\mathrm{d} #1}{\mathrm{d} #2}}

\newcommand{\id}{\mathrm{d}}
\newcommand{\mi}{\mathrm{i}}
\newcommand{\me}{\mathrm{e}}

\newcommand{\rin}{r_{\mathrm{in}}}
\newcommand{\rout}{r_{\mathrm{out}}}
\newcommand{\re}{r_{\mathrm{edge}}}

\newcommand{\Mp}{M_{\rm p}}

\newcommand{\ap}{a_{\rm p}}

\newcommand{\ep}{e_{\rm p}}

\newcommand{\Op}{\Omega_{\rm p}}
\newcommand{\Tp}{T_{\rm p}}

\newcommand{\Ms}{M_*}
\newcommand{\Mj}{M_{\rm J}}

\newcommand{\cs}{c_{\rm s}}

\newcommand{\Ok}{\Omega_{\rm K}}
\newcommand{\Mdot}{\dot{M}_0}

\date{Accepted XXX. Received YYY; in original form ZZZ}

\pubyear{2019}

\begin{document}
\label{firstpage}
\pagerange{\pageref{firstpage}--\pageref{lastpage}}
\maketitle

\begin{abstract}
We present 2D hydrodynamical simulations of hot Jupiters orbiting near the inner edge of protoplanetary discs. We systemically explore
how the accretion rate at the inner disc edge is regulated by a giant
planet of different mass, orbital separation and eccentricity.  We find
that a massive (with planet-to-star mass ratio $\gtrsim 0.003$)
eccentric ($e_p\gtrsim 0.1$) planet drives a pulsed accretion at the
inner edge of the disc, modulated at one or two
times the planet's orbital frequency. The amplitude of accretion
variability generally increases with the planet mass and eccentricity,
although some non-monotonic dependences are also possible. Applying
our simulation results to the T Tauri system CI Tau, where a young hot
Jupiter candidate has been detected, we show that the observed
luminosity variability in this system can be explained by pulsed
accretion driven by an eccentric giant planet.

\end{abstract}

\begin{keywords}
accretion,
accretion discs -- hydrodynamics -- planet-disc interactions -- planetary systems:
protoplanetary discs
\end{keywords}

\section{Introduction}
\label{sec:intro}

The detection of planets around young stars offers a direct way of testing theories of planet formation and migration. Observations of such planets remain difficult because of the strong stellar activity of their parent (T Tauri) stars, and the presence of an optically-thick disc, which can cause additional variabilities. In this context, the young ($\sim2~{\rm Myr}$) system CI Tau is of particular interest. Recently, \citet{johnskrull16} reported radial velocity detection of a short-period companion to the star. With a projected mass of $M_{\rm p}\sin I\simeq 8 \Mj$ and a period of 9 days, it represents a rare example of young hot Jupiters co-existing with protoplanetary discs. Moreover, the radial velocity data from \citet{johnskrull16} suggests a non-zero eccentricity for this object. Further measurements by \citet{flagg19} gave a planetary eccentricity of $0.25\pm0.16$, and a mass of $11.6\pm^{\tiny 2.9}_{\tiny 2.7}~\Mj$.
In addition \citet{biddle18} reported photometric variability of the star with a 9 day period, while also recovering a 6.6 day signal, presumably corresponding to the stellar rotation. The authors suggested that although the planetary companion is likely not transiting, it can still cause photometric variability by driving pulsed accretion of disc material onto the star. This is the scenario that we will examine in this paper.

The origin of the significant eccentricity of CI Tau b could be found in planet-disc interactions. A large body of hydrodynamical simulations have explored the possibility of eccentricity excitation of planets embedded discs \citep[see, e.g.,][]{papaloizou01,dlb06,rice08}. More recently, long-term simulations by \citet{rbctfm16} and \citet{ragusa18}, especially tailored for the CI Tau system, have found that the growth of planetary eccentricity is indeed possible, albeit on time-scales longer than what was previously suggested. On the theoretical side, it has long been argued that disc-planet interactions can result in a growth of eccentricity under the action of Lindblad resonances \citep{gt80,gs03,ol03,to16}. In particular, the linear analysis of \citet{to16} showed that Lindblad resonances can overcome the damping effect of corotation resonances and viscosity, and allow for the growth of eccentricity of both the disc and the planet. One important numerical result of planet-disc interactions that remains poorly understood is the fact that planets on circular orbits seem to only excite eccentricity in the disc when the planet-to-star mass ratio exceeds $\sim 0.003$ \citep[see, e.g.][]{kd06,regaly10,to17,muley19}. 

The photometric variability observed by \citet{biddle18} can also find its origin in planet-disc interactions. The variability of mass accretion in the presence of a companion can strongly differs from that of a disc around a single star. In the case of circumbinary accretion, both numerical simulations \citep[e.g.,][]{mm08,dorazio13,ml16,mml17,mml19} and observations \citep[e.g.,][]{jensen07,muzerolle13,bary14,tofflemire17,biddle18} show that accretion is variable on time-scales comparable with the binary orbital period. One interesting finding is that while circular binaries (with mass ratio of order unity) lead to a pulsed accretion at 5 times the binary period $T_b$, eccentric binaries lead to pulsed accretion with a dominant period of $T_b$ \citep[][]{ml16,mml19}. Because observations suggest that the companion to CI Tau is on an eccentric orbit, it is necessary to address the issue of pulsed accretion driven by eccentric short-period planets. In this paper we explore, by mean of hydrodynamical simulations, the accretion of disc material onto a star in the presence of an eccentric planetary companion.

This paper is organized as follows: We detail the problem setup and numerical methods in Section \ref{sec:methods}, and introduce some necessary diagnostic tools in Section \ref{sec:diagnostic}. We present results relevant for CI Tau in Section \ref{sec:resq10}, and explore a larger parameter space in Section \ref{sec:resparam}. Finally we discuss the implications of our results and conclude in Section \ref{sec:summary}. 

\section{Problem setup and numerical methods}
\label{sec:methods}

The problem of interest consists of a viscous hydrodynamics accretion disc around a star (mass $\Ms$), which is also orbited by a short-period giant planet (mass $\Mp$). We denote $q$ the planet-to-star mass ratio, and $\ap$, $\Op$, $\Tp$ and $\ep$ the semi-major axis, Keplerian mean motion, orbital period and eccentricity of the planet, respectively. The disc radial extend goes from $\rin$ to $\rout$.

We assume that the disc is locally isothermal with a pressure scale height $H$, and a constant aspect ratio $h=H/r$. Our fiducial setup uses $h=0.05$. The relation between the vertically-integrated pressure $P$ and surface density $\Sigma$ is therefore $P=\cs^2\Sigma$, where $\cs=H\Ok$ is the sound speed, and $\Ok$ the Keplerian frequency. We parametrize the kinematics viscosity through  the $\alpha$-prescription, i.e., $\nu=\alpha \cs H$; we use $\alpha=0.1$ throughout the paper.
The code units are chosen such that $G(\Ms+\Mp)=\ap=\Op=1$. Unless mentioned otherwise, we set $\rin/\ap=0.65$ and $\rout/\ap=10$.

Our initial and boundary conditions are similar to those of \citet{mml17}. The initial disc surface density is initially set to be:
\begin{equation}
\Sigma(r)=\Sigma_0 \left(\frac{r}{\ap} \right)^{-1/2}\left[1-\left(\frac{\rin}{r}\right)^{1/2} \right] \exp\left[-\left(\frac{r}{\re}\right)^{-2} \right].
\end{equation}
The exponential factor creates a cavity, speeding up the process of the cavity opening by the planet, and therefore allowing for a quicker relaxation. We chose $\re=2\ap$. 

The initial radial fluid velocity is
\begin{equation}
u_{\rm r} = -\frac{3}{2} \frac{\nu}{r} \left[1-\left(\frac{\rin}{r}\right)^{1/2} \right] ^{-1},
\end{equation}
while the azimuthal velocity is
\begin{equation}
u_\phi = r\, \Omega.
\end{equation}
The initial orbital frequency $\Omega$ is set by the balance between the pressure force and gravity, including the quadrupole force from the planet: 
\begin{equation}
\Omega^2(r) = \Ok^2 \left[1 + \frac{3}{4}\frac{q}{(1+q)^2}\left(1+\frac{3}{2}\ep^2 \right)\left(\frac{r}{\ap}\right)^{-2} \right] + \frac{1}{r\Sigma}\dd{P}{r}.
\end{equation}

At the inner edge, we use the same ``diode'' boundary condition as \citet{mml17}, which allows mass to leave the domain at the inner edge, but prevents it from re-entering. This is done by applying a zero-gradient condition ($\partial u_r/ \partial r=0$) on $u_r$ whenever it is negative, but a reflective condition on $u_r$ when it is positive. Zero-gradient conditions are imposed on $\Sigma$ and $u_\phi$. These boundary conditions mimic the loss of mass at the inner edge as the gas is accreted onto the star at the magnetospheric radius.

At the outer disc edge, we assume that mass is injected into the domain at a constant rate $\Mdot$. We apply a wave-killing zone that extend from $(\rout-1)$ to $\rout$ \citep[see, e.g.,][]{dvb06}. In this zone, all fluid quantities (generically denoted by $X$) are relaxed towards their initial value $X_0$ via:
\begin{equation}
\dd{X}{t}=-\frac{X-X_0}{t_{\rm damp}}R(r).
\end{equation}
Here $t_{\rm damp}$ is a damping time-scale, which we take to be the orbital period at the outer edge, and $R$ is a quadratic function that increase from 0 at $(\rout-1)$ to 1 at $\rout$. In code units, we chose $\Sigma_0=1$, and therefore we have $\Mdot=3\pi\alpha h^2$. 

The simulations are carried out on a grid centred on the centre of mass of the star-planet system. The planet is kept on a fixed eccentric orbit with $\ap=1$. A fluid element with coordinates $(r,\phi$) feels the total potential $\Phi$ given by
\begin{align}
\Phi(r,\phi) &= -\frac{G\Ms}{\left[r^2+r_*^2-2rr_* \cos(\phi-\phi_*) \right]^{1/2}} \nonumber \\
&-\frac{G\Mp}{\left[r^2+r_p^2-2rr_p \cos(\phi-\phi_p)+\epsilon^2\right]^{1/2}} ,
\end{align}
where $(r_*,\phi_*)$ and $(r_p,\phi_p)$ are the coordinates of the star and planet, and $\epsilon$ is a smoothing length. We take $\epsilon=0.6 H$. The star and planet do not feel the potential from the disc, and we compute their positions at each timestep by solving Kepler's equation, following the method outlined in \citet{md99}.

Finally, we follow \citet{kley99} and allow mass to accrete onto the planet. At each time-step $\Delta t$, the gas density inside the Roche lobe of the planet (radius $=\ap(q/3)^{1/3}$) is reduced by a factor $(1-0.5\Op\Delta t) $. Note that the ``accreted'' mass is not added to the mass of the planet, but simply removed from the domain. Finally, the mass of the planet is smoothly increased from 0 to $\Mp$ over the first $10\Tp$ of the simulation.

We use the {\sc pluto} code \citep{mignone12}, with a two-dimensional cylindrical grid. We use the \textit{hll} solver with a linear reconstruction method and a second-order Runge-Kutta time-integration scheme. Our fiducial grid has a resolution of $384\times 886$ in radius and azimuth, respectively. As we use a logarithmic grid in radius, this resolution ensures nearly square cells with a constant aspect ratio across the computation domain. Additional runs carried out at a higher resolution show no significant differences in accretion variabilities at the inner disc edge compared to the ones carried out at our fiducial resolution \cite[see][for more details on convergence tests for simulations with similar setup]{to17}.

\section{Diagnostic tools}
\label{sec:diagnostic}

\begin{figure}
    \begin{center}
    \includegraphics[scale=0.8]{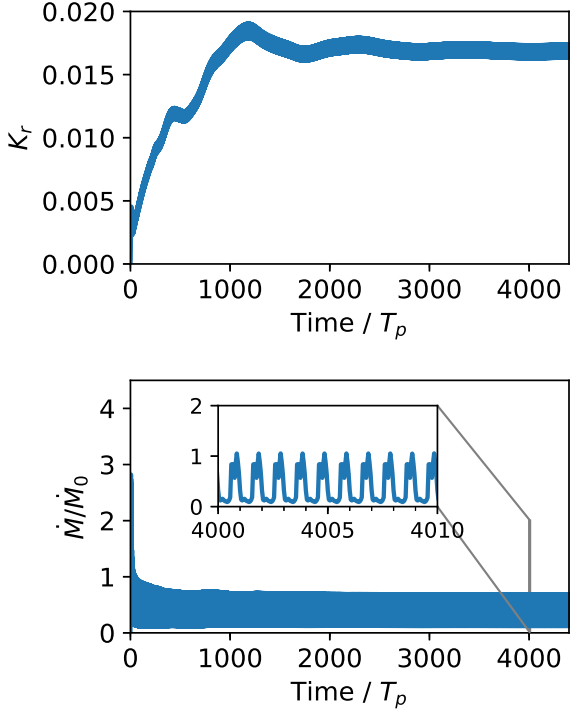}
    \caption{Time evolution of the total radial kinetic energy of the disc (top panel, in units of the Keplerian energy at $r=1$) and accretion rate at the inner edge (bottom panel).}
    \label{fig:ekr_mdot_q10}
    \end{center}
\end{figure}

The goal of this paper is to determine the accretion variability at the inner edge of the disc in the presence of a massive planetary companion. It is important to insure that the disc has reached a quasi-steady state (independent of the initial conditions). To this end, we first need to insure viscous relaxation of the inner disc region. The viscous time-scale at radius $r$ is $t_\nu=(4/9)(r^2/\nu)$. Thus after a time $t$, the disc is viscously relaxed up to a radius $r_{\rm relax}$ given by
\begin{equation}
r_{\rm relax}(t)=\left(\frac{9}{4}\alpha h^2 \Op t \right)^{2/3}\ap.
\end{equation}
With $\alpha=0.1$ and $h=0.05$, after $4000$ orbits, the disc is viscously relaxed up to $r\sim 6\ap$, which is adequate for the purpose of this study.

Secondly, pulsed accretion onto the star is driven by eccentric motion. It is therefore important to insure that the disc's eccentricity has reached a quasi-steady state. The simulations of \citet{kd06} and \citet{to17} have shown that for mass ratios of $q \gtrsim 0.003$, the outer disc would develop a significant eccentricity, which will eventually saturate. A useful quantity to measure eccentricity growth and saturation is the radial kinetic energy:
\begin{equation}
K_r=\int_{\rin}^{\rout}\int_0^{2\pi}\frac{1}{2}\Sigma u_r^2 \id\phi r\id r.
\end{equation}

Figure \ref{fig:ekr_mdot_q10} shows an example of the time evolution of the radial kinetic energy $K_r$ and the accretion rate at the disc inner edge. The latter is computed from 
\begin{equation}
\dot{M}(r,t)=-\int_0^{2\pi}r\Sigma u_r \id\phi.
\end{equation}
In practice we use a sampling of 20 measurements of the accretion rate per orbit of the planet. We see that after 4000 orbits, neither $K_r$ nor $\dot M$ show long-term variation, and the system has reached quasi-steady state. Of course, variabilities of $K_r$ and $\dot M$ on orbital timescales still exist, and we study these variabilities in the following sections. 
Note that although the simulations have reached a quasi-steady state, the time-averaged accretion rate at the inner edge is not equal to the mass supply rate at the outer edge, $\Mdot$. This is because some of the mass is ``accreted'' onto the planet and removed from the computational domain.

\section{Results for $\lowercase{q}=0.01$}
\label{sec:resq10}

\begin{figure*}
    \begin{center}
    \includegraphics[scale=0.8]{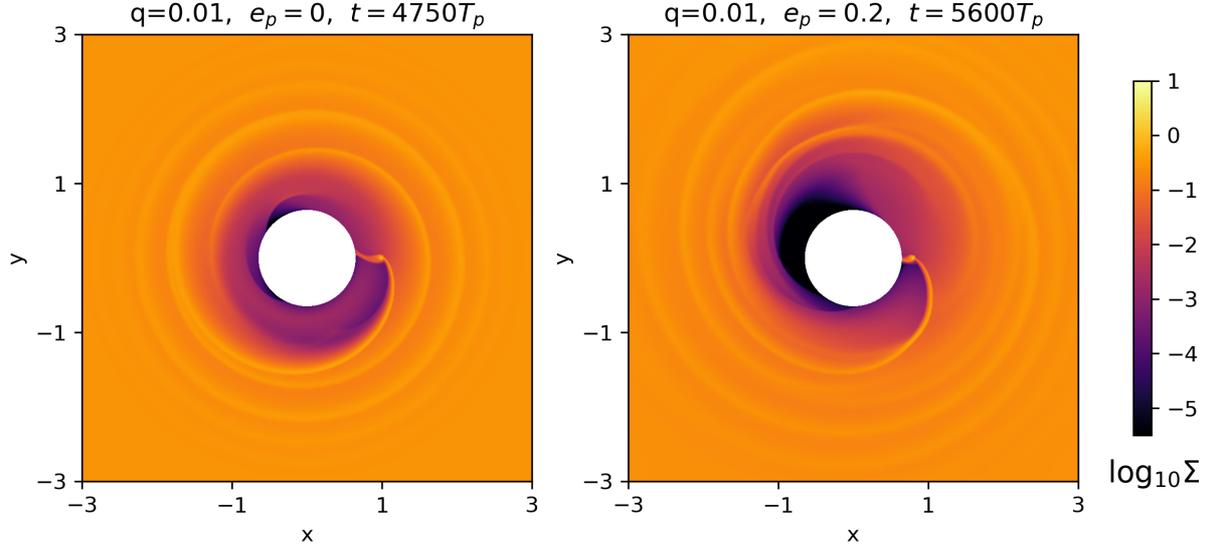}
    \caption{Disc surface density (zoomed-in to the inner part of the disc) for a planet-to-star mass ratio of $q=0.01$ and two different planet eccentricities, $\ep=0$ (left) and $\ep=0.2$ (right). In the right panel, the planet is at the pericentre.} 
    \label{fig:density_snaps}
    \end{center}
\end{figure*}

To illustrate the accretion onto the inner edge, we begin with a fiducial example with $q=0.01$, and varying the planet eccentricity. This setup is similar to the observed properties of the candidate planet around CI Tau. A snapshot of the disc surface density after several thousand orbits is shown in Figure \ref{fig:density_snaps}, where the planet is at pericentre, for two different planet eccentricities: $\ep=0$ and $\ep=0.2$. The main morphological difference is that the eccentric planet carves out a large eccentric cavity, which does not appear when the planet is on a circular orbit.

\begin{figure}
    \begin{center}
    \includegraphics[scale=0.45]{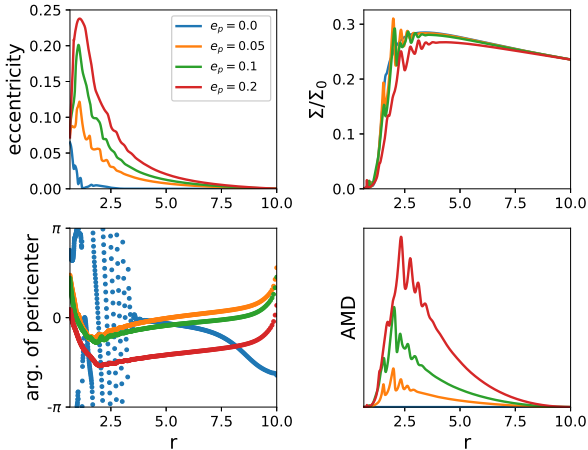}
    \caption{Azimuthally averaged radial profiles for disc eccentricity, surface density, argument of pericentre, and angular momentum deficit for different planet eccentricities.} 
    \label{fig:esa_q10}
    \end{center}
\end{figure}

This feature appears even more clearly in Figure \ref{fig:esa_q10}, where we show azimuthally averaged profiles for the disc eccentricity, surface density, argument of pericentre and angular momentum deficit (AMD). These quantities were calculated using azimuthally averaged components of the eccentricity vector of each grid cells, following \citet{to17}. In the case where $\ep=0$, the outer disc ($r>\ap$) is not eccentric; the inner disc appears to have some residual eccentricity, but since it is depleted in material, its effect is negligible (as can be seen from the fact that the AMD is nearly zero everywhere in the disc). For higher planet eccentricities, the outer disc becomes clearly eccentric, and the cavity becomes larger. The outer part of the disc also precesses nearly-rigidly. 

\begin{figure*}
    \begin{center}
    \includegraphics[scale=0.8]{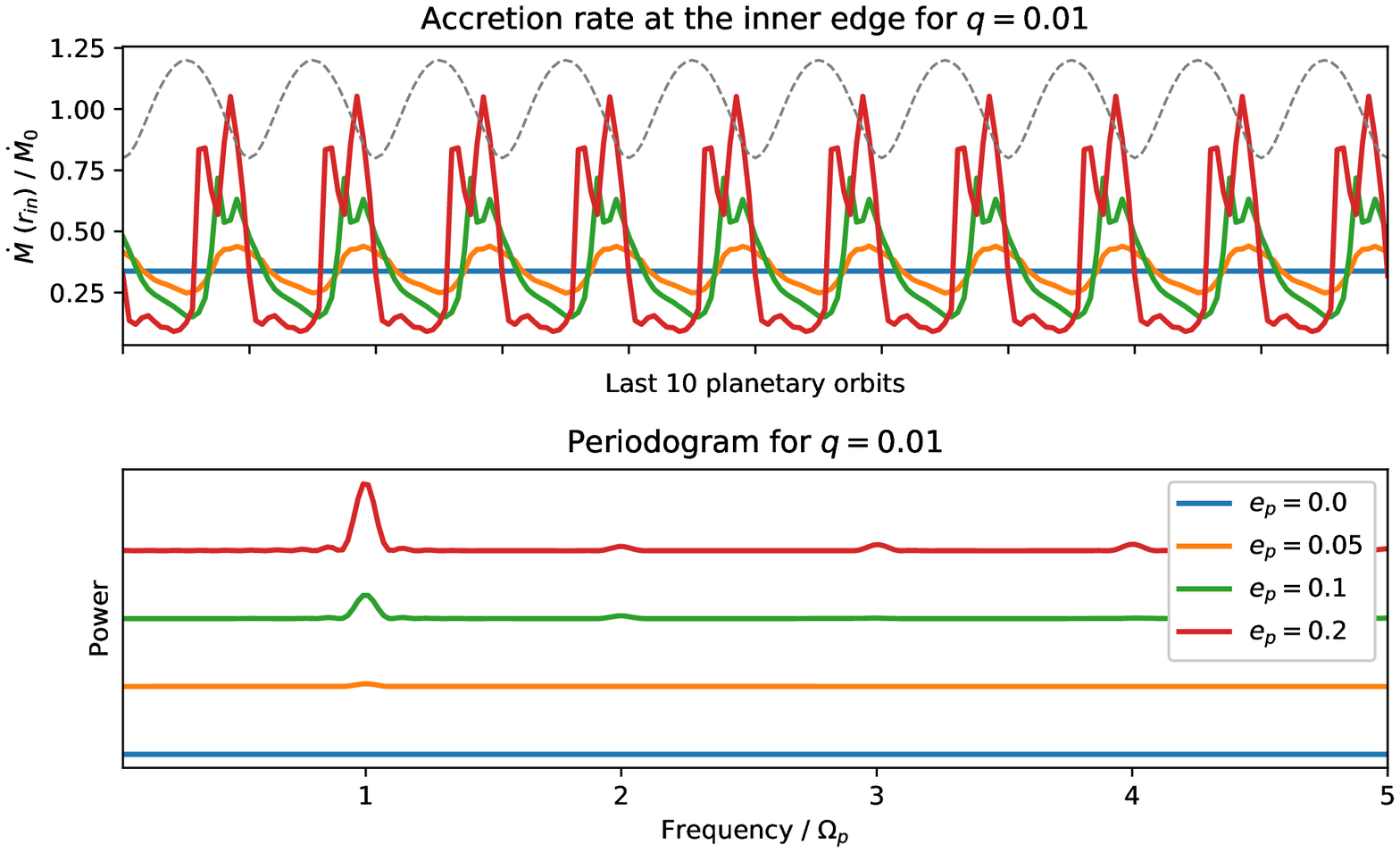}
    \caption{\textit{Top:} Accretion rate at the inner disc radius over the last 10 planetary orbits of the simulations, for $q=0.01$ and different planetary eccentricities. For comparison, the grey dashed curved shows the radial position of the planet over time. \textit{Bottom:} the corresponding periodogram. The amplitudes of the periodograms are normalized so that the amplitude of the largest signal ($\ep=0.2$) is 1, and all the other ones are scaled by the squared amplitude of the accretion rates. They are arbitrarily offset from each others for clarity.} 
    \label{fig:mdot_per_q10}
    \end{center}
\end{figure*}

\begin{figure}
    \begin{center}
    \includegraphics[scale=0.8]{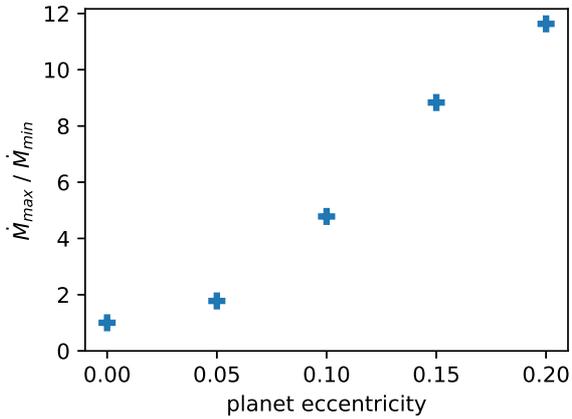}
    \caption{Peak-to-trough amplitude of the accretion rate for different eccentricities of the planet. The corresponding accretion rate time series are shown in Figure \ref{fig:mdot_per_q10}.} 
    \label{fig:max_q10}
    \end{center}
\end{figure}

The accretion rate at the inner edge, and the corresponding periodogram are shown in Figure \ref{fig:mdot_per_q10} for various planet eccentricities. As the planet eccentricity increases, the modulated accretion rate becomes more  pronounced, with a period that matches the planet's orbital period. This is confirmed when taking the Lomb-Scargle periodogram of the time-varying accretion rate. The periodogram clearly shows a main peak at the planet's orbital frequency. The amplitude of the accretion rate modulation increases with eccentricity for the range of eccentricities we consider, as shown in Figure \ref{fig:max_q10}.

\begin{figure*}
    \begin{center}
    \includegraphics[scale=0.75]{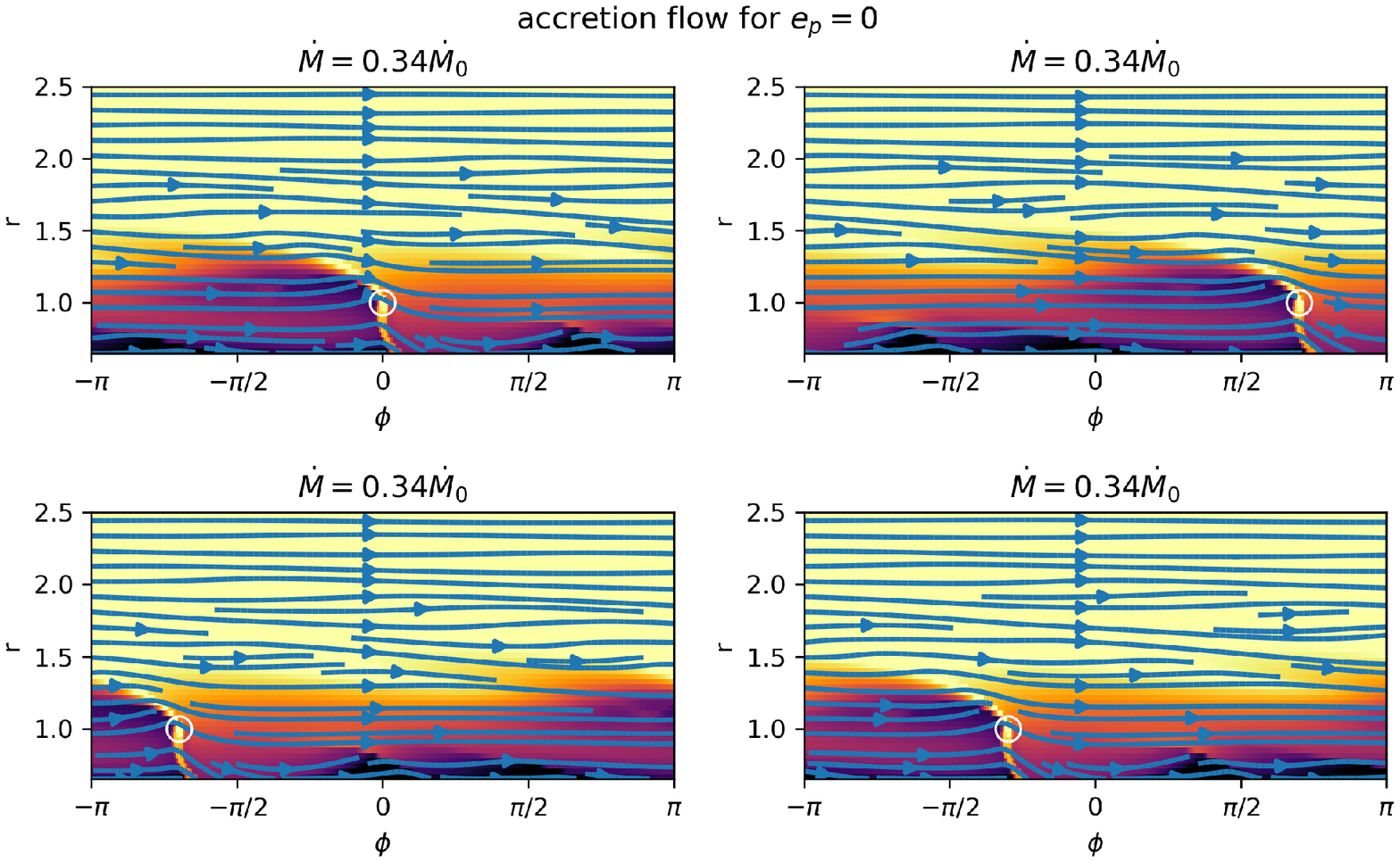}
    \caption{Streamlines overplotted on top of the surface density, for a planet-to-star mass ratio of $q=0.01$ and planet eccentricity $\ep=0$. We show four different snapshots taken over a single orbit of the planet, at different orbital phases of the planet. The position of the planet is highlighted by the white circle. The planet is at pericentre at $\phi=0$, and at apocentre at $\phi=\pi$. The corresponding $\dot{M}$ at the inner disc edge is indicated on top of each panel. In this case, the planet on a circular orbit does not cause any time variation of the accretion rate at the inner disc edge.} 
    \label{fig:flow_q10e0}
    \end{center}
\end{figure*}

\begin{figure*}
    \begin{center}
    \includegraphics[scale=0.75]{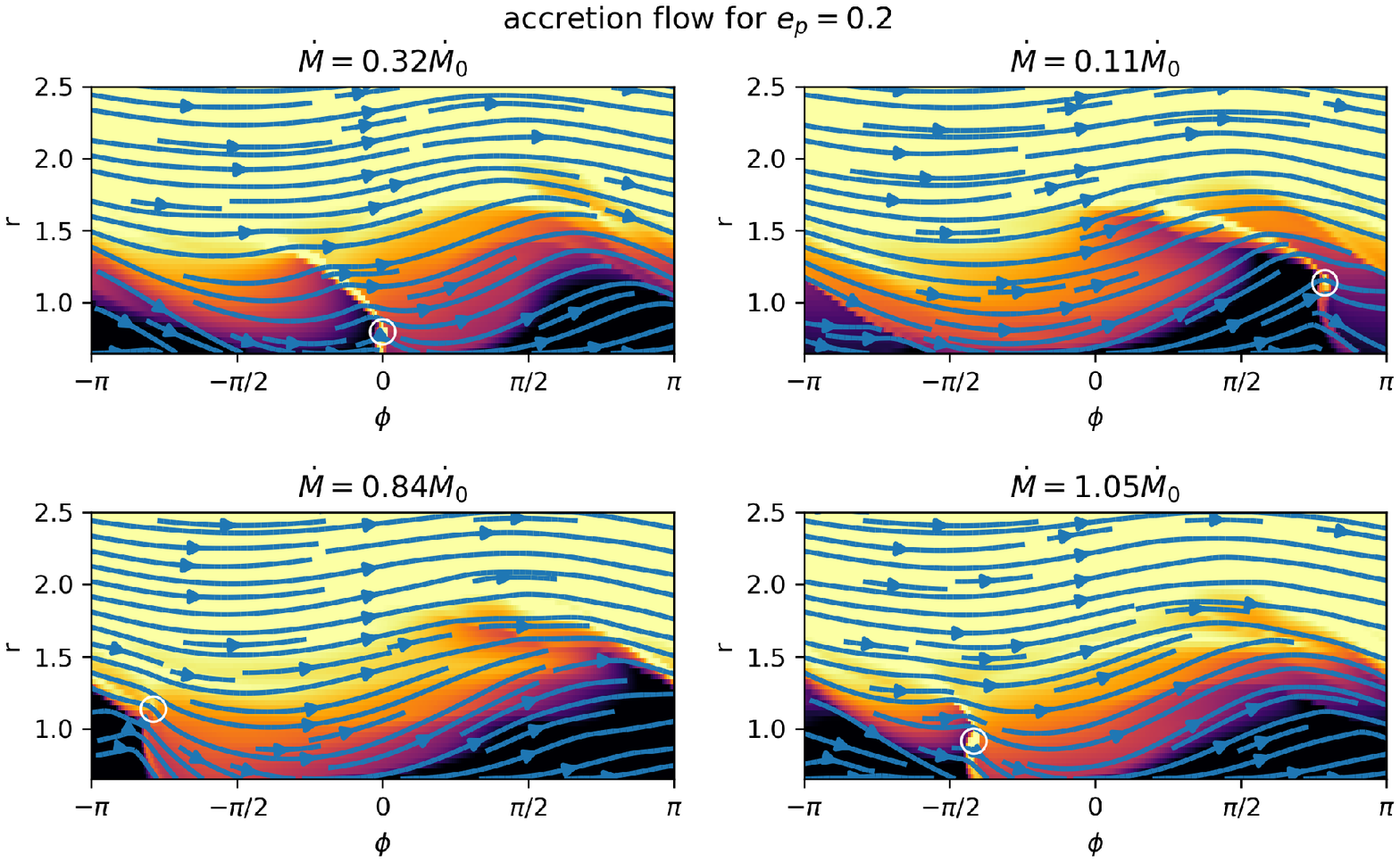}
    \caption{Same as Figure \ref{fig:flow_q10e0}, but with a planetary eccentricity of $\ep=0.2$. In this case, the morphology of the flow and the accretion rate at the disc inner edge vary over the planet's orbit.} 
    \label{fig:flow_q10e2}
    \end{center}
\end{figure*}

In Figure \ref{fig:flow_q10e0} and \ref{fig:flow_q10e2} we show the accretion flow morphology for the circular and $\ep=0.2$ cases, at different phases of the planet's orbital trajectory. The circular case exhibits a constant flow morphology at all phases, thus a flat accretion curve in Figure \ref{fig:mdot_per_q10}. However, when the planet is eccentric, the accretion rate clearly varies as the planet goes over its orbit. The two bottom panels of Figure \ref{fig:flow_q10e2} show the two episodes of maximum accretion observed in Figure \ref{fig:mdot_per_q10}: A first burst occurs when the planet enters a region of higher density and launches material inwards, closely followed by a second burst as the planet approaches the pericentre.

%From this set of simulations, the following picture emerges: when eccentric, the giant planet is capable of exciting eccentricity in the disc, and drives a pulsed accretion onto the star, with a period matching the orbital period of the planet. 
From this set of simulations, the following picture emerges: when eccentric, the giant planet drives a pulsed accretion onto the star, with a period matching the orbital period of the planet; the amplitude of the pulsation depends on the eccentricity. It is worth pointing out that in several previous works, planets on fixed circular orbits with $q=0.01$ were able to excite eccentricity in the outer disc \citep[see][]{kd06,regaly10,to17,muley19}, a feature not observed in our simulations. This is likely caused by the high viscosity used in our simulations, which results in a strong eccentricity damping. Only when the planet is eccentric can it force some eccentricity in the disc and overcome viscous damping.

\section{Exploring the parameter space}
\label{sec:resparam}

\subsection{Planet mass}
Although our study was motivated by the photometric variability observed in CI Tau, it is useful to consider other planet masses. We therefore carry out simulations for planet-to-star mass ratios ranging from $q=0.001$ to $q=0.01$. 

\begin{figure*}
    \begin{center}
    \includegraphics[scale=0.8]{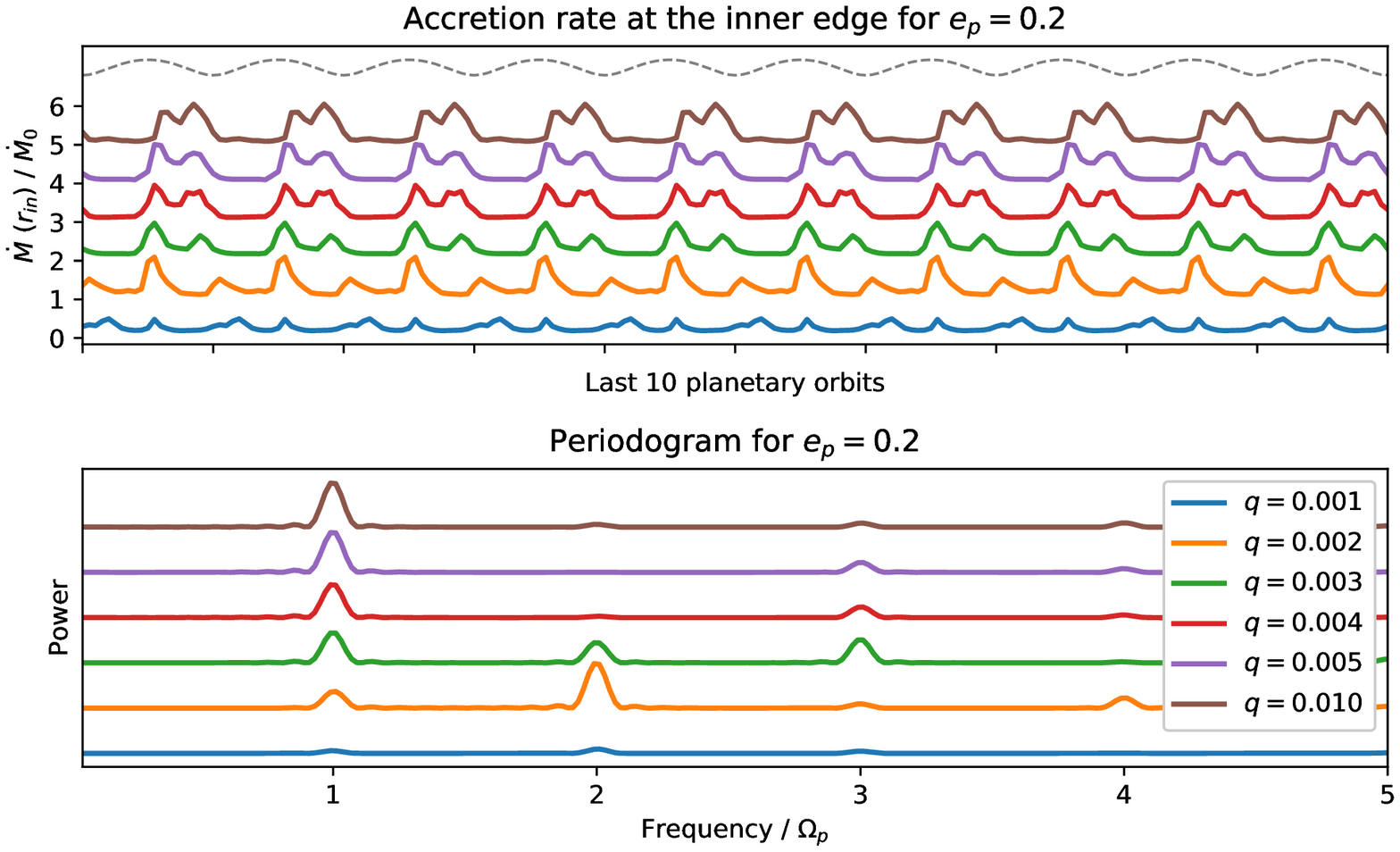}
    \caption{Same as Fig.~\ref{fig:mdot_per_q10}, but for $\ep=0.2$ and different planet-to-star mass ratio (as indicated). In the top panel, the grey dashed curve (which indicates the radial position of the planet) is offset and in arbitrary units. All the curves are offset for clarity.}
    \label{fig:mdot_per_qe2} 
    \end{center}
\end{figure*}

\begin{figure}
    \begin{center}
    \includegraphics[scale=0.8]{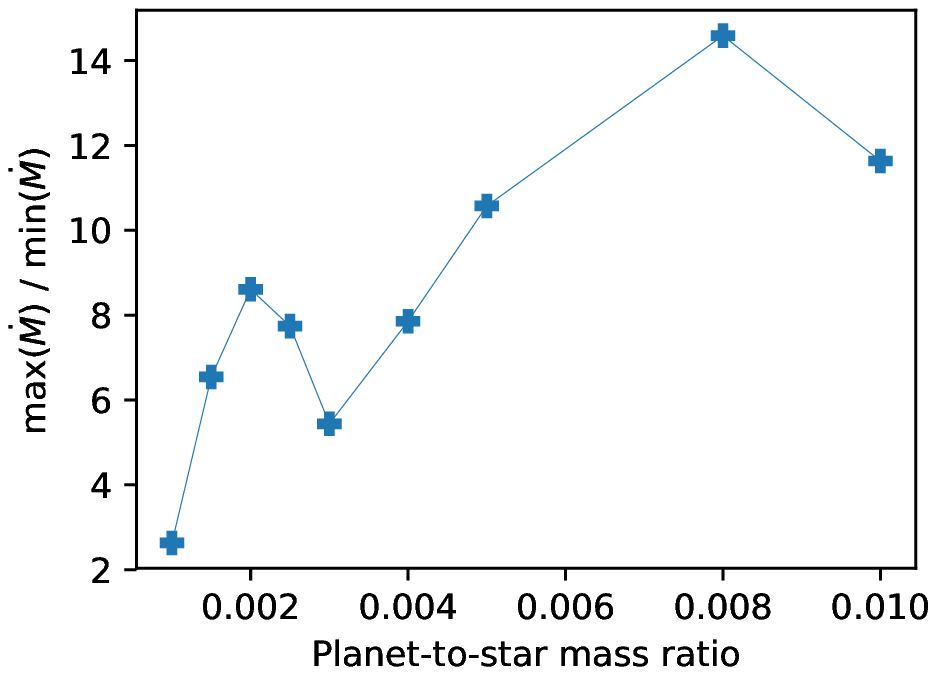}
    \caption{Peak-to-trough amplitude of the accretion rate for different planet-to-star mass ratios. The corresponding accretion rate time series are  shown in Figure \ref{fig:mdot_per_qe2}.} 
    \label{fig:max_qe2}
    \end{center}
\end{figure}

In Figure \ref{fig:mdot_per_qe2} we show the accretion rate evolution and periodogram for a planet with eccentricity $\ep=0.2$ and various planet-to-star mass ratios. For $q=0.001$, the accretion rate at the inner disc edge does not exhibit significant variations compared to the larger mass ratio 	cases. Interestingly, for $q=0.002$, large-amplitude variability at twice the planet's orbital frequency appears. As the planet mass increases, the two pulses per orbit gradually merge, and pulsed accretion at the planet's orbital frequency occurs. To illustrate this further, we plot in Figure \ref{fig:max_qe2} the peak-to-trough amplitude of the accretion rate time series shown in Figure \ref{fig:mdot_per_qe2}, with additional data points to resolve the transition around $q=0.002$. We have also added a point for $q=0.008$, where the accretion rate variation amplitude has the largest amplitude of all our simulations. This is possibly because as the planet becomes too massive, it starts accreting more mass than it lets go through the boundary, hence reducing the accretion rate variation at the inner edge. 

We have also run simulations (not shown here) with $\ep=0$ for two additional mass ratios, $q=0.005$ and $q=0.001$. We find that the accretion rate is constant over time, without any large-amplitude modulations, similar to what is shown in Figure \ref{fig:mdot_per_q10} for the case of $q=0.01$ and $\ep=0$. 

\subsection{Separation between the planet's orbit and the disc's inner edge}
\label{sec:ap}
\begin{figure*}
    \begin{center}
    \includegraphics[scale=0.65]{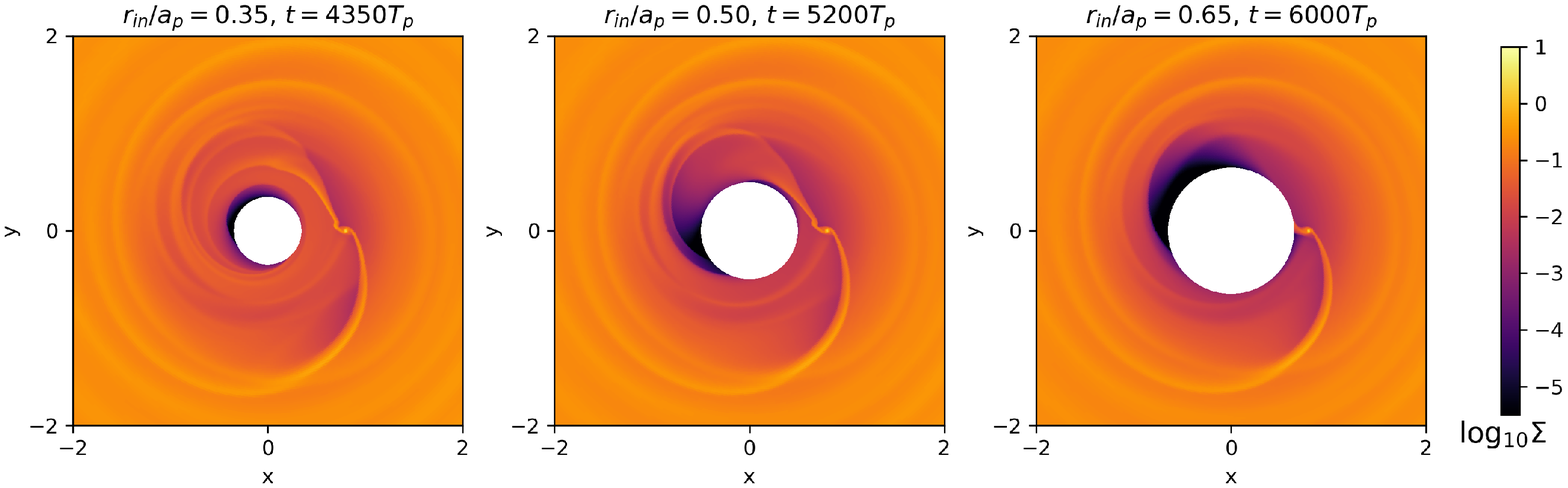}
    \caption{Disc surface density snapshots for a planet-to-star mass ratio of $q=0.004$, planet eccentricity $\ep=0.2$, and three different disc inner edge locations: $\rin/\ap=0.35$ (left), $\rin/\ap=0.5$ (centre), and $\rin/\ap=0.65$ (right).} 
    \label{fig:density_snaps_q4a}
    \end{center}
\end{figure*}

We now examine the effect of placing the planet further away from the inner edge. We consider the case of $q=0.004$ and $\ep=0.2$. In addition to our fiducial run with $\rin/\ap=0.65$, we have carried out two simulations with $\rin/\ap=0.35$ and $\rin/\ap=0.5$. Figure \ref{fig:density_snaps_q4a} shows snapshots of the disc surface density after several thousand orbits for the three cases. Because of the high viscosity, the planet does not open a clear gap, and a lot of material is able to accrete onto the inner disc edge. The corresponding periodograms show a transition from pulsed accretion at the planet's orbital frequency (when the planet is close to the edge) to twice its orbital frequency (when the planet is far for the edge). Interestingly, the variation amplitude is smaller for $\rin/\ap=0.5$ than for $\rin/\ap=0.35$ and 0.65.

\begin{figure*}
    \begin{center}
    \includegraphics[scale=0.8]{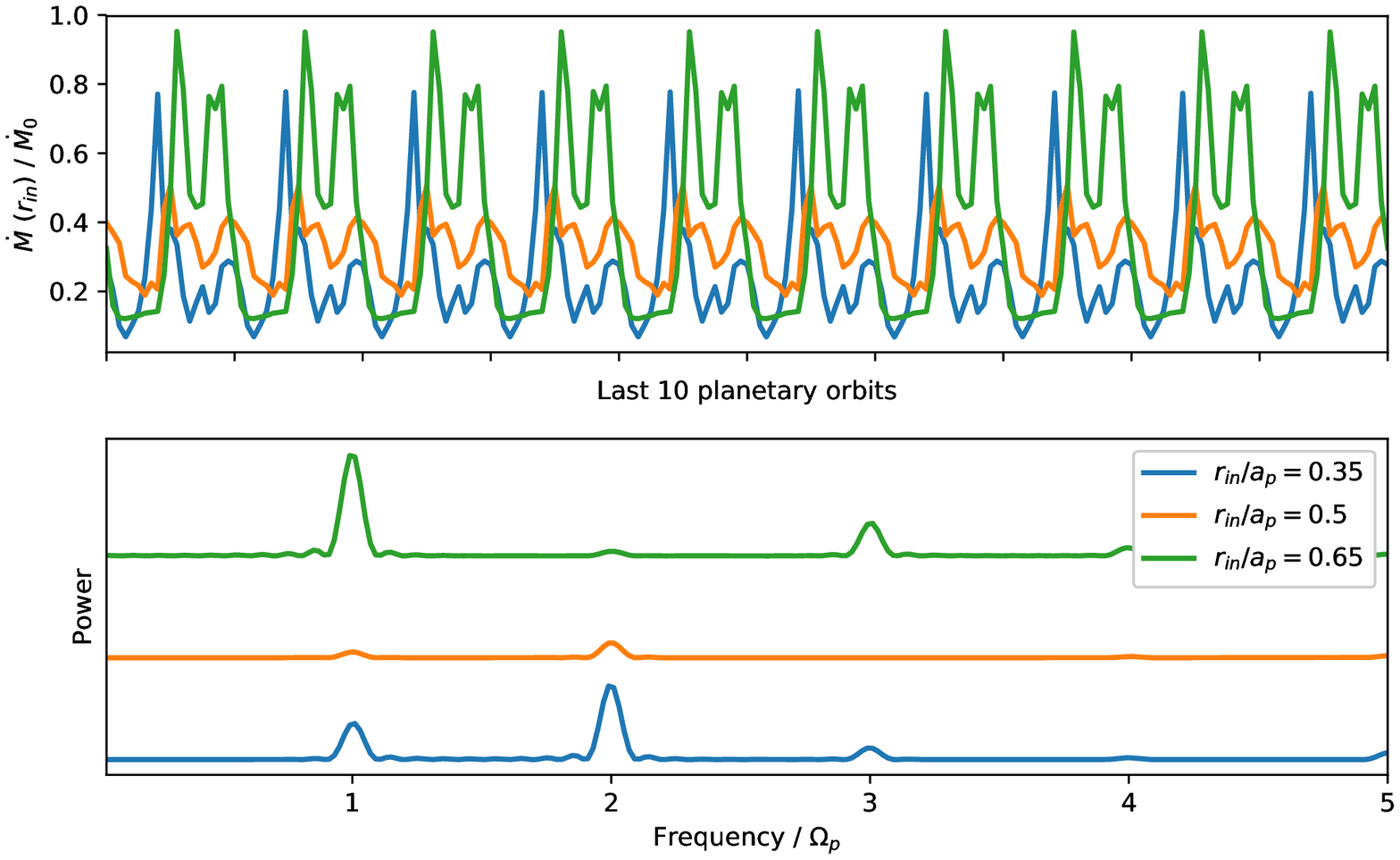}
    \caption{Same as Fig.~\ref{fig:mdot_per_q10}, except for $q=0.004$ and $\ep=0.2$, with $\rin/\ap=0.35$ (blue), $\rin/\ap=0.5$ (orange), and $\rin/\ap=0.65$ (green).}
    \label{fig:mdot_per_q4a}
    \end{center}
\end{figure*}	

\begin{figure}
    \begin{center}
    \includegraphics[scale=0.8]{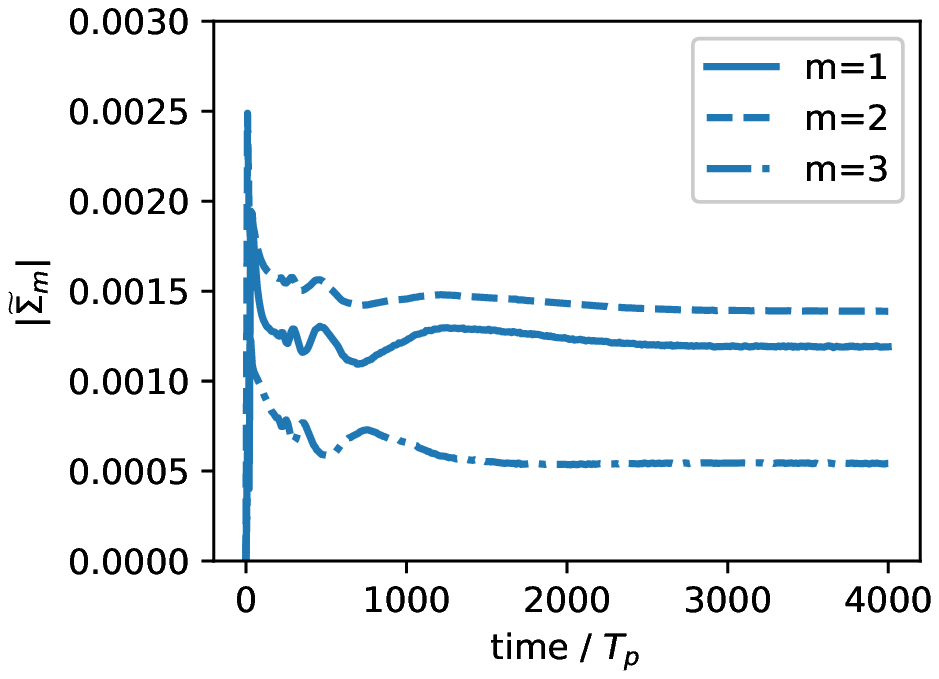}
    \caption{Time evolution of the first three Fourier components of the inner disc surface density for the case of $\rin/\ap=0.35$ shown in Figure \ref{fig:density_snaps_q4a}.} 
    \label{fig:fourier}
    \end{center}
\end{figure}

In order to understand why the $\rin/\ap=0.35$ case is dominated by the variability component at twice the planet's orbital frequency, it is useful to consider time evolution of the Fourier components of the surface density, integrated over the inner parts of the disc (between $\rin$ and $\ap$):
\begin{equation}
\tilde\Sigma_m(t)=\frac{1}{2\pi(\ap-\rin)}\int_{\rin}^{\ap}\int_0^{2\pi}\Sigma(r,\phi,t)~\me^{-\mi m \phi} \id\phi\id r.
\end{equation}
In Figure \ref{fig:fourier} we plot $\tilde\Sigma_m$ for $m=1$, 2 \& 3, showing that the $m=2$ component indeed dominates the inner disc. We speculate that this mode is dominant for $\rin/\ap=0.35$ because it corresponds to the density wave launched at the 3:1 inner Lindblad resonance, which is an $m=2$ mode excited at $\rin/\ap=0.48$ \citep{papaloizou01}. 

This feature may have important observational consequences, as it suggests that observing pulsed accretion alone is not enough to deduce the orbital period of the planet. It is necessary to break this degeneracy by other means, for instance using radial velocity measurements as was done for CI Tau.

\subsection{Planet inside the cavity}
\label{sec:inside}
So far we have considered the case where the planet's orbit lies outside the inner disc edge, i.e. $\ap (1-\ep)> \rin$.  This is consistent with the observed stellar rotation period (6 days) for CI Tau if we assume that the inner disc is truncated at the corotation radius. Indeed, if $\rin$ is determined by magnetosphere truncation, then the disc rotation rate at $\rin$ is expected to be close to the stellar rotation rate \citep[e.g.,][]{lai14a}.

However, it is possible that the inner disc radius of CI Tau is larger than the corotation radius. Indeed modelling of the CI Tau disc by \citet{mcclure13} has suggested an inner disc edge at 0.12 au, which would put CI Tau b inside the inner cavity of the disc. More generally, it is useful to consider the case of a planet orbiting inside the inner edge of the disc; such a case may occur at the end-stage of disc migration, as disc torques may push the planet slightly inwards of the disc inner truncation radius.

We therefore ran additional simulations with $\ap(1+\ep)<\rin$. The numerical setup is the same, the only difference being that the planet is no longer in the computational domain. We still observed the pulsed accretion phenomenon. The accretion rate time series and the corresponding periodograms look similar to the case where the planet orbits in the disc. Since these plots are similar to Fig.~\ref{fig:mdot_per_q10}, we simply summarize our findings in Table \ref{tab:inside}, which lists the ratio $\dot{M}_{\max}/\dot{M}_{\min}$ (similar to what is plotted in Figures \ref{fig:max_q10} and \ref{fig:max_qe2}), as well as the dominant frequency of the pulsed accretion. We see that larger mass planets lead to larger amplitude pulsed accretion (for $q=0.001$, the amplitude of the pulse is almost zero, regardless of whether the planet is eccentric or not). For planets on circular orbits we did not observe any pulsed accretion. In almost all cases where pulsed accretion is observed, the dominant frequency of the pulsation is the planet's orbital frequency. The only exception is the case $q=0.001$, $\rin/\ap=1.5$ and $\ep=0.2$, where two peaks of equal amplitude are observed at $\Op$ and $2\Op$, although in this case the amplitude of variation in the accretion rate is very small.

\begin{table}
	\centering
	\caption{Results of simulations for a planet inside the disc's inner edge. For each planet-to-star mass ratio $q$, separation between planet and disc inner edge $\rin/\ap$, and planet eccentricity $\ep$, we give the peak-to-trough amplitude of the accretion rate at the inner edge $\dot{M}_{\min}/\dot{M}_{\min}$, and the dominant frequency of the pulsed accretion (in units of the planet's orbital frequency).}
	\label{tab:inside}
	\begin{tabular}{lcccc} 
		\hline
		$q$ & $\rin/\ap$ & $\ep$ & $\dot{M}_{\max}/\dot{M}_{\min}$ & Dominant frequency ($\Op$) \\
		\hline
		0.010 & 1.35 & 0.2 & 3.68 & 1 \\
		0.010 & 1.50 & 0.2 & 2.54 & 1 \\
		0.010 & 1.65 & 0.2 & 2.39 & 1 \\
		0.004 & 1.35 & 0.2 & 1.40 & 1 \\
		0.004 & 1.50 & 0.2 & 1.40 & 1 \\
		0.004 & 1.65 & 0.2 & 1.39 & 1 \\
		0.001 & 1.35 & 0.2 & 1.13 & 1 \\
		0.001 & 1.50 & 0.2 & 1.06 & 1--2 \\
		0.001 & 1.65 & 0.2 & 1.06 & 1 \\
		0.001 & 1.35 & 0.0 & 1.00 & -- \\
		0.001 & 1.50 & 0.0 & 1.00 & -- \\
		0.001 & 1.65 & 0.0 & 1.00 & -- \\
		\hline
	\end{tabular}
\end{table}

\section{Summary and Discussion}
\label{sec:summary}

In this paper we have used hydrodynamical simulations to study the
accretion variabilities at the inner edge of a protoplanetary disc in
the presence of a hot Jupiter. This is motivated by the recent
observations of the T-Tauri star CI Tau with a young hot Jupiter
candidate. We have shown that a giant planetary companion
(planet-to-star mass ratio of $\sim 0.005$) on an eccentric orbit can
drive pulsed accretion onto the star with a period matching the
orbital period of the planet, in agreement with the observation. We
have also found that an equally massive but circular planet would not
produce such a large-amplitude pulsed accretion.  This suggest that
the candidate giant planet around CI Tau is indeed eccentric.

More generally, we have systematically investigated how the accretion variabilities
depend on the planet mass, orbital semi-major axis and eccentricity.
For the typical disc parameters adopted in this paper (aspect ratio $H/r=0.05$ 
and viscosity $\alpha=0.1$), we find the following trends: 

(1) An eccentric planet can drive large-amplitude pulsed accretion onto the
central star, with the dominant variability frequency given by one or
two times the planet's orbital frequency $\Omega_p$.  The variability amplitude
generally increases with the planet's eccentricity.  A circular giant
planet produces a small or negligible accretion variability (see
Figs.~\ref{fig:mdot_per_q10}-\ref{fig:max_q10}).

(2) More massive planets tend to drive accretion variability with large amplitudes, 
although the dependence may not be strictly monotonic (see Figs.~\ref{fig:mdot_per_qe2}-\ref{fig:max_qe2}).

(3) The accretion variability depends in a non-trivial way on the planet's semi-major axis
relative to the inner disc radius (Sections \ref{sec:ap} and \ref{sec:inside}). For a given planetary mass
and eccentricity, the dominant variability frequency may change from $\Omega_p$ to
$2\Omega_p$ and the amplitude may change non-monotonically as $\rin/\ap$ changes
(see Fig.~\ref{fig:mdot_per_q4a}).

We emphasize that some of our more quantitative results (such as the
non-monotonic trends) must be considered tentative. An important
limitation of the work reported here is that our simulations only
covered the region outside the inner disc truncation radius ($\rin$). 
We used an inflow ``diode'' boundary condition at $r=\rin$ 
(see Section 2), and this boundary condition cannot be justified
rigorously without actually simulating the flow inside $\rin$.
A real protoplanetary disc is likely truncated
by the magnetic field of the central star. So an ``authentic'' numerical 
study of the accretion variability induced by a planet would involve full 3D MHD 
simulations of magnetosphere accretion, a complicated 
task that is beyond the scope of the paper
\citep[see][for a review]{lai14a,romanova14}. In any case, the long-term 
simulations required to attain ``quasi-steady'' results, as well as the coverage of 
a large parameter space, as we have attempted in this
paper using viscous hydrodynamics, would be difficult to replicate 
in full MHD simulations. Overall, despite the caveats of the simulations reported in 
this paper, we believe that our general results and trends are valid, 
at least qualitatively \citep[see][for a similar problem 
of circumbinary accretion]{mml17,mml19}. These results can serve as a guide for ongoing 
observations of young hot Jupiters in protoplanetary discs and for future
more sophisticated numerical simulations.

\section*{Acknowledgements}
This work has been supported in part by NASA grants NNX14AG94G and 80NSSC19K0444, and NSF grant AST-1715246. The research done in this project
made use of the SciPy stack \citep{scipy}, including
NumPy \citep{numpy} and Matplotlib \citep{matplotlib}, as well as Astropy,\footnote{http://www.astropy.org} a community-developed core Python package for Astronomy \citep{astropy:2013, astropy:2018}.

\bibliographystyle{mnras}
\bibliography{biblio2}

\begin{thebibliography}{}
\makeatletter
\relax
\def\mn@urlcharsother{\let\do\@makeother \do\$\do\&\do\#\do\^\do\_\do\%\do\~}
\def\mn@doi{\begingroup\mn@urlcharsother \@ifnextchar [ {\mn@doi@}
  {\mn@doi@[]}}
\def\mn@doi@[#1]#2{\def\@tempa{#1}\ifx\@tempa\@empty \href
  {http://dx.doi.org/#2} {doi:#2}\else \href {http://dx.doi.org/#2} {#1}\fi
  \endgroup}
\def\mn@eprint#1#2{\mn@eprint@#1:#2::\@nil}
\def\mn@eprint@arXiv#1{\href {http://arxiv.org/abs/#1} {{\tt arXiv:#1}}}
\def\mn@eprint@dblp#1{\href {http://dblp.uni-trier.de/rec/bibtex/#1.xml}
  {dblp:#1}}
\def\mn@eprint@#1:#2:#3:#4\@nil{\def\@tempa {#1}\def\@tempb {#2}\def\@tempc
  {#3}\ifx \@tempc \@empty \let \@tempc \@tempb \let \@tempb \@tempa \fi \ifx
  \@tempb \@empty \def\@tempb {arXiv}\fi \@ifundefined
  {mn@eprint@\@tempb}{\@tempb:\@tempc}{\expandafter \expandafter \csname
  mn@eprint@\@tempb\endcsname \expandafter{\@tempc}}}

\bibitem[\protect\citeauthoryear{{Astropy Collaboration} et~al.,}{{Astropy
  Collaboration} et~al.}{2013}]{astropy:2013}
{Astropy Collaboration} et~al., 2013, \mn@doi [\aap]
  {10.1051/0004-6361/201322068}, \href
  {http://adsabs.harvard.edu/abs/2013A%26A...558A..33A} {558, A33}

\bibitem[\protect\citeauthoryear{{Bary} \& {Petersen}}{{Bary} \&
  {Petersen}}{2014}]{bary14}
{Bary} J.~S.,  {Petersen} M.~S.,  2014, \mn@doi [\apj]
  {10.1088/0004-637X/792/1/64}, \href
  {https://ui.adsabs.harvard.edu/abs/2014ApJ...792...64B} {792, 64}

\bibitem[\protect\citeauthoryear{{Biddle}, {Johns-Krull}, {Llama}, {Prato}  \&
  {Skiff}}{{Biddle} et~al.}{2018}]{biddle18}
{Biddle} L.~I.,  {Johns-Krull} C.~M.,  {Llama} J.,  {Prato} L.,   {Skiff}
  B.~A.,  2018, \mn@doi [\apjl] {10.3847/2041-8213/aaa897}, \href
  {http://adsabs.harvard.edu/abs/2018ApJ...853L..34B} {853, L34}

\bibitem[\protect\citeauthoryear{{D'Angelo}, {Lubow}  \& {Bate}}{{D'Angelo}
  et~al.}{2006}]{dlb06}
{D'Angelo} G.,  {Lubow} S.~H.,   {Bate} M.~R.,  2006, \mn@doi [\apj]
  {10.1086/508451}, \href {http://adsabs.harvard.edu/abs/2006ApJ...652.1698D}
  {652, 1698}

\bibitem[\protect\citeauthoryear{{D'Orazio}, {Haiman}  \&
  {MacFadyen}}{{D'Orazio} et~al.}{2013}]{dorazio13}
{D'Orazio} D.~J.,  {Haiman} Z.,   {MacFadyen} A.,  2013, \mn@doi [\mnras]
  {10.1093/mnras/stt1787}, \href
  {http://adsabs.harvard.edu/abs/2013MNRAS.436.2997D} {436, 2997}

\bibitem[\protect\citeauthoryear{{Flagg}, {Johns-Krull}, {Nofi}, {Llama},
  {Prato}, {Sullivan}, {Jaffe}  \& {Mace}}{{Flagg} et~al.}{2019}]{flagg19}
{Flagg} L.,  {Johns-Krull} C.~M.,  {Nofi} L.,  {Llama} J.,  {Prato} L.,
  {Sullivan} K.,  {Jaffe} D.~T.,   {Mace} G.,  2019, \mn@doi [\apj]
  {10.3847/2041-8213/ab276d}, \href
  {https://ui.adsabs.harvard.edu/abs/2019ApJ...878L..37F} {878, L37}

\bibitem[\protect\citeauthoryear{{Goldreich} \& {Sari}}{{Goldreich} \&
  {Sari}}{2003}]{gs03}
{Goldreich} P.,  {Sari} R.,  2003, \mn@doi [\apj] {10.1086/346202}, \href
  {http://adsabs.harvard.edu/abs/2003ApJ...585.1024G} {585, 1024}

\bibitem[\protect\citeauthoryear{{Goldreich} \& {Tremaine}}{{Goldreich} \&
  {Tremaine}}{1980}]{gt80}
{Goldreich} P.,  {Tremaine} S.,  1980, \mn@doi [\apj] {10.1086/158356}, \href
  {http://adsabs.harvard.edu/abs/1980ApJ...241..425G} {241, 425}

\bibitem[\protect\citeauthoryear{Hunter}{Hunter}{2007}]{matplotlib}
Hunter J.~D.,  2007, \mn@doi [Computing in Science \& Engineering]
  {10.1109/MCSE.2007.55}, 9, 90

\bibitem[\protect\citeauthoryear{{Jensen}, {Dhital}, {Stassun}, {Patience},
  {Herbst}, {Walter}, {Simon}  \& {Basri}}{{Jensen} et~al.}{2007}]{jensen07}
{Jensen} E. L.~N.,  {Dhital} S.,  {Stassun} K.~G.,  {Patience} J.,  {Herbst}
  W.,  {Walter} F.~M.,  {Simon} M.,   {Basri} G.,  2007, \mn@doi [\aj]
  {10.1086/518408}, \href
  {https://ui.adsabs.harvard.edu/abs/2007AJ....134..241J} {134, 241}

\bibitem[\protect\citeauthoryear{{Johns-Krull} et~al.,}{{Johns-Krull}
  et~al.}{2016}]{johnskrull16}
{Johns-Krull} C.~M.,  et~al., 2016, \mn@doi [\apj]
  {10.3847/0004-637X/826/2/206}, \href
  {http://adsabs.harvard.edu/abs/2016ApJ...826..206J} {826, 206}

\bibitem[\protect\citeauthoryear{{Kley}}{{Kley}}{1999}]{kley99}
{Kley} W.,  1999, \mn@doi [\mnras] {10.1046/j.1365-8711.1999.02198.x}, \href
  {http://adsabs.harvard.edu/abs/1999MNRAS.303..696K} {303, 696}

\bibitem[\protect\citeauthoryear{{Kley} \& {Dirksen}}{{Kley} \&
  {Dirksen}}{2006}]{kd06}
{Kley} W.,  {Dirksen} G.,  2006, \mn@doi [\aap] {10.1051/0004-6361:20053914},
  \href {http://adsabs.harvard.edu/abs/2006A%26A...447..369K} {447, 369}

\bibitem[\protect\citeauthoryear{{Lai}}{{Lai}}{2014}]{lai14a}
{Lai} D.,  2014, in European Physical Journal Web of Conferences. p. 01001
  (\mn@eprint {arXiv} {1402.1903}), \mn@doi{10.1051/epjconf/20136401001}

\bibitem[\protect\citeauthoryear{{MacFadyen} \&
  {Milosavljevi{\'c}}}{{MacFadyen} \& {Milosavljevi{\'c}}}{2008}]{mm08}
{MacFadyen} A.~I.,  {Milosavljevi{\'c}} M.,  2008, \mn@doi [\apj]
  {10.1086/523869}, \href {http://adsabs.harvard.edu/abs/2008ApJ...672...83M}
  {672, 83}

\bibitem[\protect\citeauthoryear{{McClure} et~al.,}{{McClure}
  et~al.}{2013}]{mcclure13}
{McClure} M.~K.,  et~al., 2013, \mn@doi [\apj] {10.1088/0004-637X/775/2/114},
  \href {https://ui.adsabs.harvard.edu/abs/2013ApJ...775..114M} {775, 114}

\bibitem[\protect\citeauthoryear{{Mignone}, {Zanni}, {Tzeferacos}, {van
  Straalen}, {Colella}  \& {Bodo}}{{Mignone} et~al.}{2012}]{mignone12}
{Mignone} A.,  {Zanni} C.,  {Tzeferacos} P.,  {van Straalen} B.,  {Colella} P.,
    {Bodo} G.,  2012, \mn@doi [\apjs] {10.1088/0067-0049/198/1/7}, \href
  {http://adsabs.harvard.edu/abs/2012ApJS..198....7M} {198, 7}

\bibitem[\protect\citeauthoryear{{Miranda}, {Mu{\~n}oz}  \& {Lai}}{{Miranda}
  et~al.}{2017}]{mml17}
{Miranda} R.,  {Mu{\~n}oz} D.~J.,   {Lai} D.,  2017, \mn@doi [\mnras]
  {10.1093/mnras/stw3189}, \href
  {http://adsabs.harvard.edu/abs/2017MNRAS.466.1170M} {466, 1170}

\bibitem[\protect\citeauthoryear{{Mu{\~n}oz} \& {Lai}}{{Mu{\~n}oz} \&
  {Lai}}{2016}]{ml16}
{Mu{\~n}oz} D.~J.,  {Lai} D.,  2016, \mn@doi [\apj]
  {10.3847/0004-637X/827/1/43}, \href
  {http://adsabs.harvard.edu/abs/2016ApJ...827...43M} {827, 43}

\bibitem[\protect\citeauthoryear{{Mu{\~n}oz}, {Miranda}  \& {Lai}}{{Mu{\~n}oz}
  et~al.}{2019}]{mml19}
{Mu{\~n}oz} D.~J.,  {Miranda} R.,   {Lai} D.,  2019, \mn@doi [\apj]
  {10.3847/1538-4357/aaf867}, \href
  {https://ui.adsabs.harvard.edu/abs/2019ApJ...871...84M} {871, 84}

\bibitem[\protect\citeauthoryear{{Muley}, {Fung}  \& {van der Marel}}{{Muley}
  et~al.}{2019}]{muley19}
{Muley} D.,  {Fung} J.,   {van der Marel} N.,  2019, arXiv e-prints, \href
  {https://ui.adsabs.harvard.edu/abs/2019arXiv190207191M} {p. arXiv:1902.07191}

\bibitem[\protect\citeauthoryear{{Murray} \& {Dermott}}{{Murray} \&
  {Dermott}}{1999}]{md99}
{Murray} C.~D.,  {Dermott} S.~F.,  1999, {Solar system dynamics}.
Cambridge University Press

\bibitem[\protect\citeauthoryear{{Muzerolle}, {Furlan}, {Flaherty}, {Balog}  \&
  {Gutermuth}}{{Muzerolle} et~al.}{2013}]{muzerolle13}
{Muzerolle} J.,  {Furlan} E.,  {Flaherty} K.,  {Balog} Z.,   {Gutermuth} R.,
  2013, \mn@doi [\nat] {10.1038/nature11746}, \href
  {https://ui.adsabs.harvard.edu/abs/2013Natur.493..378M} {493, 378}

\bibitem[\protect\citeauthoryear{{Ogilvie} \& {Lubow}}{{Ogilvie} \&
  {Lubow}}{2003}]{ol03}
{Ogilvie} G.~I.,  {Lubow} S.~H.,  2003, \mn@doi [\apj] {10.1086/368178}, \href
  {http://adsabs.harvard.edu/abs/2003ApJ...587..398O} {587, 398}

\bibitem[\protect\citeauthoryear{Oliphant}{Oliphant}{06  }]{numpy}
Oliphant T.,  2006--, {NumPy}: A guide to {NumPy}, USA: Trelgol Publishing,
  \url {http://www.numpy.org/}

\bibitem[\protect\citeauthoryear{{Papaloizou}, {Nelson}  \&
  {Masset}}{{Papaloizou} et~al.}{2001}]{papaloizou01}
{Papaloizou} J.~C.~B.,  {Nelson} R.~P.,   {Masset} F.,  2001, \mn@doi [\aap]
  {10.1051/0004-6361:20000011}, \href
  {http://adsabs.harvard.edu/abs/2001A%26A...366..263P} {366, 263}

\bibitem[\protect\citeauthoryear{{Price-Whelan} et~al.,}{{Price-Whelan}
  et~al.}{2018}]{astropy:2018}
{Price-Whelan} A.~M.,  et~al., 2018, \mn@doi [\aj] {10.3847/1538-3881/aabc4f},
  \href {https://ui.adsabs.harvard.edu/#abs/2018AJ....156..123T} {156, 123}

\bibitem[\protect\citeauthoryear{{Ragusa}, {Rosotti}, {Teyssandier}, {Booth},
  {Clarke}  \& {Lodato}}{{Ragusa} et~al.}{2018}]{ragusa18}
{Ragusa} E.,  {Rosotti} G.,  {Teyssandier} J.,  {Booth} R.,  {Clarke} C.~J.,
  {Lodato} G.,  2018, \mn@doi [\mnras] {10.1093/mnras/stx3094}, \href
  {http://adsabs.harvard.edu/abs/2018MNRAS.474.4460R} {474, 4460}

\bibitem[\protect\citeauthoryear{{Reg{\'a}ly}, {S{\'a}ndor}, {Dullemond}  \&
  {van Boekel}}{{Reg{\'a}ly} et~al.}{2010}]{regaly10}
{Reg{\'a}ly} Z.,  {S{\'a}ndor} Z.,  {Dullemond} C.~P.,   {van Boekel} R.,
  2010, \mn@doi [\aap] {10.1051/0004-6361/201014427}, \href
  {http://adsabs.harvard.edu/abs/2010A%26A...523A..69R} {523, A69}

\bibitem[\protect\citeauthoryear{{Rice}, {Armitage}  \& {Hogg}}{{Rice}
  et~al.}{2008}]{rice08}
{Rice} W.~K.~M.,  {Armitage} P.~J.,   {Hogg} D.~F.,  2008, \mn@doi [\mnras]
  {10.1111/j.1365-2966.2007.12817.x}, \href
  {http://adsabs.harvard.edu/abs/2008MNRAS.384.1242R} {384, 1242}

\bibitem[\protect\citeauthoryear{{Romanova}, {Lovelace}, {Bachetti}, {Blinova},
  {Koldoba}, {Kurosawa}, {Lii}  \& {Ustyugova}}{{Romanova}
  et~al.}{2014}]{romanova14}
{Romanova} M.~M.,  {Lovelace} R.~V.~E.,  {Bachetti} M.,  {Blinova} A.~A.,
  {Koldoba} A.~V.,  {Kurosawa} R.,  {Lii} P.~S.,   {Ustyugova} G.~V.,  2014, in
  European Physical Journal Web of Conferences. p. 05001 (\mn@eprint {arXiv}
  {1311.4597}), \mn@doi{10.1051/epjconf/20136405001}

\bibitem[\protect\citeauthoryear{{Rosotti}, {Booth}, {Clarke}, {Teyssandier},
  {Facchini}  \& {Mustill}}{{Rosotti} et~al.}{2017}]{rbctfm16}
{Rosotti} G.~P.,  {Booth} R.~A.,  {Clarke} C.~J.,  {Teyssandier} J.,
  {Facchini} S.,   {Mustill} A.~J.,  2017, \mn@doi [\mnras]
  {10.1093/mnrasl/slw184}, \href
  {http://adsabs.harvard.edu/abs/2017MNRAS.464L.114R} {464, L114}

\bibitem[\protect\citeauthoryear{{Teyssandier} \& {Ogilvie}}{{Teyssandier} \&
  {Ogilvie}}{2016}]{to16}
{Teyssandier} J.,  {Ogilvie} G.~I.,  2016, \mn@doi [\mnras]
  {10.1093/mnras/stw521}, \href
  {http://adsabs.harvard.edu/abs/2016MNRAS.458.3221T} {458, 3221}

\bibitem[\protect\citeauthoryear{{Teyssandier} \& {Ogilvie}}{{Teyssandier} \&
  {Ogilvie}}{2017}]{to17}
{Teyssandier} J.,  {Ogilvie} G.~I.,  2017, \mn@doi [\mnras]
  {10.1093/mnras/stx426}, \href
  {http://adsabs.harvard.edu/abs/2017MNRAS.467.4577T} {467, 4577}

\bibitem[\protect\citeauthoryear{{Tofflemire}, {Mathieu}, {Herczeg}, {Akeson}
  \& {Ciardi}}{{Tofflemire} et~al.}{2017}]{tofflemire17}
{Tofflemire} B.~M.,  {Mathieu} R.~D.,  {Herczeg} G.~J.,  {Akeson} R.~L.,
  {Ciardi} D.~R.,  2017, \mn@doi [\apj] {10.3847/2041-8213/aa75cb}, \href
  {https://ui.adsabs.harvard.edu/abs/2017ApJ...842L..12T} {842, L12}

\bibitem[\protect\citeauthoryear{{Virtanen} et~al.,}{{Virtanen}
  et~al.}{2019}]{scipy}
{Virtanen} P.,  et~al., 2019, arXiv e-prints, \href
  {https://ui.adsabs.harvard.edu/abs/2019arXiv190710121V} {p. arXiv:1907.10121}

\bibitem[\protect\citeauthoryear{{de Val-Borro} et~al.,}{{de Val-Borro}
  et~al.}{2006}]{dvb06}
{de Val-Borro} M.,  et~al., 2006, \mn@doi [\mnras]
  {10.1111/j.1365-2966.2006.10488.x}, \href
  {http://adsabs.harvard.edu/abs/2006MNRAS.370..529D} {370, 529}

\makeatother
\end{thebibliography}

\label{lastpage}
\end{document}